%% file: arxiv_version.tex
\documentclass{article}

\usepackage[sfdefault,scaled=.9]{noto}
\usepackage[proportional,scaled=1]{erewhon}
\usepackage[english]{babel}
\RequirePackage[protrusion=true,expansion=true,final,babel]{microtype}

\RequirePackage{makecell}

\usepackage[font={footnotesize,sf}]{caption}

\RequirePackage{calc}
\RequirePackage{pstricks}
\RequirePackage{hyphenat}

\usepackage{tabularx}  
\usepackage{booktabs}  
\usepackage[mode=text]{siunitx} 
\usepackage{graphicx} 
\usepackage{csquotes}\MakeOuterQuote{"}  
\usepackage[hidelinks]{hyperref} 

\usepackage{subcaption}
\usepackage{multirow}

\usepackage{authblk}

\RequirePackage{csquotes}
\RequirePackage[style=apa,sortcites=true,sorting=nyt,backend=biber]{biblatex}

\DeclareLanguageMapping{american}{american-apa}

\title{Detecting Coordinated Behaviour on Video-First Platforms\\{\large The Challenge of Multimodality and Complex Similarity on TikTok}}
\author[1]{Inga K. Wohlert}
\author[1]{Davide Vega}
\author[1]{Matteo Magnani}
\author[2]{Alexandra Segerberg}
\date{}

\affil[1]{Infolab, Dept. of Information Technology, Uppsala University, Sweden}
\affil[2]{Dept. of Government, Uppsala University, Sweden}

\begin{document}
\maketitle

\abstract{Research on online coordinated behaviour has predominantly focused on text-based social media platforms.
However, the rise of video-first platforms such as TikTok introduces distinct challenges. 
The multimodal nature of video posts, combining visuals, audio, and text, allows for coordination across various modalities and complicates comparison between posts.
This paper proposes an approach to detecting coordination that addresses these characteristic challenges. Our methodology, based on multilayer network analysis, is tailored to capture coordination across multiple modalities, and explicitly handles complex forms of similarity inherent in video and audio content. We test this approach on German political posts regarding the 2024 European Elections retrieved via the TikTok Research API. Our results demonstrate the ability of our approach to identify coordination within the constraints of the API, while also critically highlighting potential pitfalls and limitations. \newline \textbf{keywords:} coordinated behaviour, video-first platform, TikTok, social media, multilayer network, multimodality, complex similarity}

\input{body}

\section{Funding}
We acknowledge that work on this research has been funded through eSSENCE, a Swedish national strategic research program in e-Science, and Swedish Research Council grant 2021-02769. Additionally, experiments were enabled by resources provided by the National Academic Infrastructure for Supercomputing in Sweden (NAISS), partially funded by the Swedish Research Council through grant agreement no. 2022-06725. 

\printbibliography

\end{document}

%% file: body.tex
\section{Introduction}\label{sec1}

Methods for detecting coordinated behaviour are increasingly important in the study of online communication \parencite{Manocci2024}. Social media platforms now reach a significant portion of the population \parencite{Merten2020, Reuters2024} and serve as important channels for news consumption \parencite{Giglietto2020, Espinoza2023, Zerback2021}. They have consequently become important arenas in which to influence public opinion. While some forms of coordination on social media are legitimate, others are manipulative and play a role in digital election interference and disinformation campaigns \parencite{Nizzoli2021, Thiele2025}. One example of manipulative coordination is astroturfing, where multiple accounts create a false appearance of widespread grassroots support for a particular cause \parencite{Kovic2018, Zerback2021}.

Existing methods for detecting coordinated behaviour in social media have focused mainly on text-based platforms such as X (formerly known as Twitter) and Facebook \parencite{Manocci2024}, overlooking the rise in popularity of video-first platforms, such as TikTok. However, the nature of video posts, combining visuals, audio, and text, poses specific challenges. 
The first challenge is multimodality. On video-first platforms, nearly all posts integrate multiple modalities,
such as video, audio, and semi-structured text.
A coordinated attempt to increase the visibility of a message can in principle target any combination of those modalities. For example, a set of coordinated videos could show different visual content, but use the same music and text to implicitly signal that these videos are part of the same political campaign, or are associated to a common ideology, or present a homogenous message. 
The second challenge is the complex similarity between posts. Assessing whether two posts enact the same message within the same modality can be ambiguous and technically demanding. For example, multiple users may repost a video with minor modifications, such as adding a few frames or changing a few spoken words in the audio, still providing a sign of potential coordination. Identifying such modifications, 
and how they impact the message of the post, is non-trivial. 

To address these challenges, we contribute a novel  
approach to detect coordination on video-first platforms. The approach has two key features. First, we address the prominence of multimedia content by accounting for coordination across
multiple modalities, represented as the layers of a multilayer network. In particular, we consider video frames, audio transcript, music, video description, URLs and hashtags, both individually and in combination. Second, we select and validate similarity functions   
to evaluate whether groups of related (but not necessarily identical) video-based social media posts are part of the same coordinated information diffusion process.

We apply our method to TikTok. TikTok is an archetypical video-first platform that is known for being specifically a recommender-based short-video system, and it is well-established as a major social media platform with political relevance. 
The platform’s user base has expanded beyond the predominantly young demographic and right-wing content of its early years \parencite{Brodovskaya2022, Weimann2020}.
Various types of actors are now well-represented on the platform. Although TikTok prohibits paid political ads, it has become an important arena for political campaigns of various kinds. Political parties in several countries conduct election campaigns on the platform  
\parencite{Cervi2021, Widholm2024, Grantham2024, tiktok_politics_religion_culture, Albertazzi2023}, which is also used for advocacy campaigns and protest mobilisation \parencite{Brodovskaya2022, Hohner2024}, as a space for playful meme politics \parencite{Vijay2021, Cervi2023, Hautea2021}, and in war propaganda \parencite{Bsch2024}. 
The relevance of TikTok for the detection of coordinated behaviour was accentuated during the recent Romanian elections, during which TikTok users were reportedly paid to promote a pro-Russian candidate who had previously received minimal recognition and subsequently performed well in the elections \parencite{tiktok_romanian_elections, ifes_romanian_election_annulment, bbc_tiktok_romanian_election}.

We test our approach on a dataset we collected based on the German politics surrounding the 2024 European Elections. Our approach yields evidence of potential coordination in this dataset. We find accounts with similar username structure that were only active during the election period. We also find that different groups of user accounts show signs of coordinated behaviour across different modalities, which confirms the need for a multimodal approach.  
Some of the evidence of potential coordination is obtained through the use of similarity functions that match related but not identical content, which highlights the importance of such similarity functions in the context of video-first environments.

\section{Detecting online coordination}

A rich literature has emerged in the study of Coordinated Behaviour and Influence (CBI) 
with particular focus on Facebook and Twitter. The term ``coordinated inauthentic behaviour'' has its origin at Facebook \parencite{facebook_inauthentic}, where it was used to refer to actors working together to mislead others. Other platforms employ different strategies and terminology. On TikTok, similar actions are tracked under the term of ``covert influence operations'' \parencite{tiktok_covert_influence_operations}.
Following \textcite{Manocci2024}, we use the term ``coordinated online behaviour'' to refer to a group of actors (e.g.~accounts, groups, pages) who perform synergic actions in pursuit of an intent. 

A key challenge is to define and detect ``signs of coordinated behaviour", since we generally cannot assess synergy and intent directly when we use social media APIs to analyse behaviour. In this section we therefore first describe a commonly used sign of coordinated behaviour, known as co-action, which has been successfully applied to text-based platforms such as Facebook and X.  Then, we summarise how previous work has explored co-actions in the context of the two challenges associated with video-first platforms: multimodality  
and complex similarity. 

\subsection{Detecting coordinated behaviour using co-actions}
\label{sec:sota_coactions}

A sign that is frequently used on platforms such as Facebook and Twitter to detect coordinated behaviour is ``co-actions''. In this context, the term co-action is typically defined as two ``actors'' performing the same ``action'' on the same ``target'', usually within the same ``timeframe''. Examples of targets are social media posts (e.g. a Tweet) and hashtags or URLs contained inside a post. Examples of actions are sharing or commenting a post or specifying a hashtag or URL inside a post \parencite{KirdemirAA22, Jahn_2023, Manocci2024, MannocciPhD}. The most studied forms of action are retweeting \parencite{Nizzoli2021, Pacheco2021} and sharing identical elements such as Web links \parencite{Giglietto2020,Gruzd2022}.
Regarding the timeframe, the typical requirement in the literature for two actions to be considered a co-action is that they occur almost simultaneously. This poses a methodological challenge to define and identify a cut-off point for what is to count as sufficiently simultaneously. Usually, researchers set a fixed timeframe, e.g.~one minute \parencite{Soares2023}, or calculate the timeframe based on the data \parencite{Giglietto2020}. 

The co-action approach to detecting coordination is promising also for studying coordinated behaviour on short-video platforms. In order to extend the approach in that direction, however, it needs traction on two characteristic features of short-video platforms: their advanced multimodality and potential for complex similarity between posts.

\subsection{Co-actions and multimodality}
\label{sec:multi}
 
A key feature of short-video platforms is that posts integrate modalities such as video, audio, and text in such an integral way that none of these modalities is likely to be sufficient on its own to understand the post but any one of the modalities can be important for detecting coordination. Yet, combining multiple modalities in one approach is still a largely unexplored topic in research building on a co-action approach: Modalities are typically considered one at a time and separately of each other \parencite{Luceri2025}. 

This said, there is a trend in the literature away from simpler methods in which only one modality is considered. Empirical studies of coordinated behaviour have recently begun to consider more than one modality. In those works, ``modality'' is broadly interpreted as one of multiple possible types of co-actions  (such as a retweet, a reply, or posting a hashtag) \parencite{Magelinski2022, DeClerck2024, Venncio2024}. Given the growing interest in considering more than one type of co-action, \textcite{MannocciPhD} performed the first methodological study of multimodal detection of coordinated behaviour, comparing the results of different analysis pipelines on the same Twitter data.
We propose to extend the focus on multiple modalities also to the different aspects of the same video-based post (such as visuals, audio, and text).

A few studies have considered coordinated behaviour across multiple social media platforms \parencite{Barbero2023, Cinus2025} and time periods \parencite{Tardelli2024}. These studies share with our work the idea of using multilayer networks in the context of coordinated behaviour detection. However, they do not focus on multimodal coordination inside one platform.

\subsection{Co-actions and complex similarity}
\label{sec:simil}
Another key feature of contemporary video-first platforms is the potential for complex similarity between posts. These online spaces support platforms cultures in which it is common for users to echo other posts with minor modifications. This complicates comparison between posts, and relying on the identification of identical content to trace coordination has less traction.

A small number of studies are forging a promising path towards similarity-based coordination detection, i.e. similarity functions that take two potential targets (e.g. video frames, audio transcripts) as input and return a value indicating how similar they are. Co-actions targeting similar but not necessarily identical content include posting tweets containing similar texts, or posting similar images or videos \parencite{Pacheco2021, DeClerck2024}. 
Some approaches convert images into text using machine learning to analyse descriptions rather than visuals \parencite{Stampe2023}, which has been shown to better capture connotative aspects of the images \parencite{Arminio25}. Others have explored comparing text within images \parencite{Soares2023} and metadata \parencite{Yu2021}. To identify similar posts, researchers create content networks (e.g.~text-text or image-image) that can be transformed into user-user coordination networks \parencite{huiXian_2022, Ng2022}. 

Similar to the present paper, recent work independently performed by \textcite{Luceri2025} applies similarity functions to TikTok posts. 
This work highlights challenges in formulating meaningful functions, especially for audio transcripts.

\section{A Method to Detect Coordination on TikTok}
\label{sec:methodology}

In this work, we propose a multilayer network-based method, informed by prior literature on CBI detection. 
In order to account for TikTok's multimodal structure, we construct a multilayer coordination network where each layer captures a distinct type of co-action (e.g.~posting the same sequence of hashtags, posting videos with very similar audio transcripts or descriptions, or posting videos with very similar visual content). This multilayered actor-centric model enables us to comprehensively analyse interactions between user accounts, simultaneously capturing signs of coordinated behaviour within individual modalities and across multiple modalities.

\begin{table}[]
    \centering
    \begin{tabular}{llp{6cm}}
        & Step & Objective \\
        \hline
        \#1 & Layer selection & Define the type of co-actions to be considered, e.g. two accounts posting equivalent videos.\\
        \#2 & Similarity thresholds & Define when two complex targets (e.g. videos, audio transcripts), not necessarily identical, should be considered equivalent, thus potentially defining a co-action.\\
        \#3 & Freq.~and temp.~thresholds & Filter some of the co-actions that are not due to coordination.\\
        \#4 & Multilayer clustering & Identify groups of potentially coordinated accounts, within or across layers/modalities.\\
        \#5 & Cluster characterisation & Exclude false positives and describe the remaining clusters using account and content information.\\
    \end{tabular}
    \caption{Objectives of the five steps in our detection methods}
    \label{tab:method}
\end{table}

The five steps of our method are summarised in Table~\ref{tab:method}, and detailed in the next subsections.
To build the multilayer network within which we look for coordinated groups of accounts (clusters), we first decide which types of co-actions we consider in the study, among those that can be extracted using the TikTok Research API. Each layer defined during the first step corresponds to a specific type of co-action.
During the second step, we fine-tune similarity functions, defining which pairs of targets constitute a co-action when targeted by two actors. We do this separately for each layer where complex similarity is involved.  
This ensures that minor yet significant variations --- such as slightly altered reposted videos --- are effectively captured as possible signs of coordination, overcoming the limitations of simpler, exact-matching methods. In the third step, each layer undergoes a customised filtering process based on empirically validated thresholds. 
The fourth step is a multilayer cluster analysis performed on the resulting multilayer network. This produces a number of clusters, some of which may correspond to coordinated groups of users. The fifth step consists in characterising these clusters, filtering those that (after a qualitative check) do not indicate coordination, and enriching the others with information from their users and the content they interacted with. 

\subsection{Step 1: Layer selection}
\label{subsec:sim}

The first step in our method involves the construction of a multilayer coordination network, similar to the approach used by \textcite{MannocciPhD} to study different types of text-based co-actions on Twitter (now X). 
On TikTok three data elements constitute a post and can be used to extract targets: a video (including visuals and audio), music, and a textual description. In addition, a post itself can be the target of some actions, in which case the post ID is the data used to define the co-actions.
Each layer is defined by a type of target and a similarity function used to decide when two targets should be considered equivalent, thus potentially being part of a co-action.

Table~\ref{table_methods} lists possible layers that can be constructed for TikTok.
From a video we can obtain a set of video frames, that can be used to inspect visual similarity, and a transcript of the audio. Music is listed separately, because TikTok assigns an ID to music elements that can be re-used in different videos. From a video description, we can compare the whole text, but we can also extract hashtags and URLs from the description as possible targets of coordinated visibility amplification. In addition, layers can be created by user interaction with the posts.

\begin{table*}[t]
\small
\centering
\begin{tabular}{llll}
\toprule
\textbf{Action} & \textbf{Target}   & \textbf{Similarity function} &\textbf{Data}\\
\toprule
Posting & Visuals  & near exact matching of frames & Video \\
Posting & Audio  & edit distance of transcript & Video \\
Posting & Audio  & partial edit distance of transcript & Video \\ 
Posting & Music  & exact matching & Music ID \\ 
Posting & Text  & exact matching & Description \\
Posting & URL  & exact matching & Description \\
Posting & Hashtag Sequence  & exact matching & Description\\
Posting & Hashtag  & exact matching & Description\\  
Liking & Post  & exact matching & Post ID \\
Commenting & Post  & exact matching & Post ID \\ 
Reposting & Post  & exact matching & Post ID \\ 
Stitch & Post  & exact matching & Post ID \\ 
Duett & Post & exact matching  & Post ID \\ 
\bottomrule
\end{tabular}
\caption{Some actions, targets, and similarity functions that can be used to construct multimodal co-action layers. We also indicate the input data from which each target is obtained.}
\label{table_methods}
\normalsize
\end{table*}

\begin{itemize}
    \item \textbf{Visual Similarity.} To detect visual coordination, we apply a perceptual image hashing method that enables approximate matching between video frames. We sample frames  and compare each frame from the shorter video to frames in the longer one, requiring that all frames have a near-equivalent match. The number and distribution of frames depends on available computational resources, e.g. all frames, or one frame every few seconds, or a frame for each scene as done by \cite{segerberg_visual_2025}. This technique is designed to tolerate minor differences while detecting visually similar content.

    \item \textbf{Same Audio.} 
    Because of minor transcription errors or subtle modifications (e.g.~trimming), two videos may have identical or nearly identical audio without sharing the same audio transcript. To capture such cases, we use edit distance metrics to measure the textual similarity of audio transcripts, quantifying how many operations are needed to transform one transcript into another. 
    
    \item \textbf{Partial Audio.} This layer uses the same target and edit distance function as Same Audio, but considers two audio transcripts to be the same target if one is a subset of the other.

    \item \textbf{Music ID.} The Music ID is an audio identifier provided via the TikTok Research API. 
    This identifier is assigned automatically either when a user reposts or reuses audio from another existing post --- typically by sharing or remixing that content --- or when the users manually assign a curated, trending audio during video creation. Thus, while it is possible for two users to independently select the same popular audio, it is equally plausible that shared Music IDs result from internal redistribution or imitation of existing content. This layer helps identify patterns of coordinated reuse or algorithmically encouraged convergence around specific audio trends.

    \item \textbf{Video Description.} Exact repetition of the same description across different accounts may indicate coordinated messaging or the use of templated content.

    \item \textbf{URL.} Video descriptions may contain URLs, which we can extract using regular expressions, to then consider identical URLs to be the same target. We note that some studies in the literature on detection of coordinated behaviour use the domain of the URLs as targets. However, two accounts posting URLs from the same domain (but not the same URL) may show some ideological alignment, but no direct sign of coordinated information spreading. In addition, very common domains (such as youtube.com) would have to be removed, because they do not imply any coordination. 

    \item \textbf{Hashtag Sequence.} 
    As for URLs, we can extract hashtags from video descriptions using regular expressions. In this case, the hashtag sequence is already provided as a list by the API. Then, we can consider identical sequences of hashtags to be the same target.

    \item \textbf{Hashtag.} 
    In the literature, individual hashtags are often used as targets, and in principle they can also be used for video-first platforms.  
    In general, the same hashtag can be seen as a weaker sign of coordination. Therefore, if individual hashtags are used to construct a layer, some popular hashtags may have to be filtered to reduce the number of false positives.
   
    \item \textbf{Like.} In a platform strongly based on algorithmic recommendation, interacting with a post can be a way to amplify its visibility. One of those interactions is likes. While TikTok allows users to share their liked videos publicly, only few users currently do so, making this form of co-action invisible within the Researcher API.

    \item \textbf{Comment.} A second form of interaction with a post is commenting. Similar to likes, the user behind comments underneath videos are kept anonymised by the Researcher API. Therefore, the action cannot be connected to a user account.

    \item \textbf{Repost.} A common way to increase the reach of a post on many platforms is through reposting. On TikTok, a repost does not create its own post, but adds the video to the repost page of the user. Therefore, information about reposts need to be collected for each user, which can be challenging depending on the number of users considered in the study.

    \item \textbf{Duet.} A different way to share a video posted by a different user is through a duet, which combines the original visuals with new visual content of the user creating the duet. However, creating a duet does not entail that the user agrees with the message of the original video. Hence, this layer may lead to false positives.

    \item \textbf{Stitch.} Finally, a user can create a stitch, which allows the user to pause and comment on the original video. Similar to the duet a stitch does not mean that the user is agreeing with the message of the original video and therefore this can lead to false positives in the layer.
\end{itemize}

\subsection{Step 2: Similarity threshold tuning}
\label{subsec:thr}

Once targets and similarity functions have been defined, thresholds have to be decided for all functions that are not based on exact matchings. Fine-tuning the similarity functions addresses the difficulty of complex similarity. For this, a set of videos should be sampled and annotated in regards to the similarity of the videos. Based on the annotations, the algorithmic thresholds are tuned to optimise precision and recall. In this way, two targets are considered to be the same only if their similarity exceeds the decided threshold. 

\subsection{Step 3: Frequency and temporal filtering}
\label{subsec:filtering}

After tuning the thresholds for the similarity functions, we also apply a filter to each layer by removing edges based on (i) the frequency of co-actions between pairs of users, and/or (ii) the temporal proximity of actions performed on the same target. The goal of this filtering process is twofold: to eliminate noisy or spurious connections and to retain structures that meaningfully reflect patterns of potential coordination. While there are fixed filtering thresholds that are commonly used on other platforms and studies, in this method the presence of multiple layers requires to identify multiple thresholds which may differ on different layers.

\subsection{Step 4: Multilayer clustering}
\label{subsec:com}

The identification of clusters of users in the multilayer coordination network can be performed using many available algorithms \parencite{magnani_community_2021}. \textcite{MannocciPhD} suggests that the choice of algorithm depends on the type of relations between social media users that one wants to identify. Following this consideration, we recommend running multiple methods, and in the empirical part of this work we use both generalised Louvain \parencite{Mucha2010} and multilayer clique percolation \parencite{AfsarmaneshTehrani2018}  as two distinct approaches --- the former being often used in the literature, the latter designed to only discover clusters where complete subgraphs above a given size are present in the coordination network. 

\subsection{Step 5: Cluster characterisation}

The objective of this step is to perform a last filter, aimed at removing or splitting the clusters that do not correspond to coordinated groups (false positives), and to provide richer descriptions of the remaining clusters based on additional information not available in the co-action networks.

Any method to detect coordinated behaviour using co-actions has to consider the difference between signs of coordination in the data and actual coordination between the actors producing the data. In practice, there is a risk that false positives arise when the same type of co-action is caused by coordinated behaviour as well as by other processes not involving any coordination. On TikTok, at least two known sources of false positives should be checked. The first is the fact that TikTok is designed to encourage a memetic communication that involves the repetition, imitation and referencing of highly similar material \parencite{Zulli2020, Cervi2023, Hautea2021}. For example, the platform urges users to add trending music to their videos. These algorithmic incentives to perform similar actions and to engage with the same popular material may lead to co-action without underlying coordination. The second concerns clusters of co-actions targeting similar but not identical content. For example, in some cases the small changes to the content of a video do not alter the original message, and may thus be part of a coordinated effort to amplify that message, but in other cases the added visual cues highlight disagreement with the underlying video, leading to very similar videos that do not necessarily constitute a sign of coordinated behaviour.

The analysis of individual clusters is also useful for characterising the type of coordination that generated them. This step may include qualitative analysis of the content, to be tailored on the data at hand. Signals that may be relevant to look at include: similar account names, synchronised posting and focused posting activity around the time of the event, the ideological nature of the content, and whether the same accounts recur in multiple clusters.

\section{Case study}

In this section, we describe an application of our method to study coordination on TikTok during the 2024 European Parliament elections in Germany. 
We present different types of clusters that we were able to identify, and show how 
considering multiple modalities and fine-tuning layer-specific similarity functions 
contributed to their detection. 

\subsection{Dataset}
\label{subsec:data}

We focus on the 2024 European Parliament election in Germany, held on June 9, 2024. We used the TikTok Research API to perform two complementary data collections, one to fetch posts produced by political parties, one to obtain posts mentioning political parties in their  hashtags. The two sets of posts were then merged and analysed together.

We began our data collection by gathering TikTok videos posted by accounts associated with the main political parties and their lead candidates at both national and state levels, during the 30-day period preceding the election (i.e.~May 10 to June 9, 2024). For this first data collection, we extended the list of accounts of the 2021 German elections study by \textcite{Righetti2022}. We then filtered the collected posts to include only those from parties that secured at least three seats in the European election, matching the minimum number of seats held by the smallest party represented in the Bundestag\footnote{The national parliament of the Federal Republic of Germany} at that time. We excluded posts from parties with minimal media presence but retained content from all represented parties in the Bundestag, as well as some newly established parties. Notable examples for newly established parties include BSW and Volt. BSW is a party established only a few months prior to the EU elections, which has since gained a significant presence in the German political sphere. Volt, a smaller party, achieved notable success among younger voters during the EU campaign by leveraging social media, particularly TikTok \parencite{tagesschau2024kleinparteien}.

In addition, we identified and selected the primary hashtag representing each party included in the first collection (\texttt{afd}, \texttt{cdu}, \texttt{bsw}, \texttt{nurmitlinks}, \texttt{freiedemokraten}, \texttt{freiewähler}, \texttt{grüne}, \texttt{spd}, \texttt{votevolt}) and we retrieved all posts containing at least one of the hashtags within the final 14 days before the election. 

The resulting dataset comprises 24,463 videos in various formats (e.g.~stand-alone, slideshows), including 959 videos originally posted by the main party and candidate accounts. For each of the posts, we gathered the original TikTok video as well as the metadata associated with it as returned by the Research API. 

\subsection{Applying the Method}

\subsubsection{Step 1: Layer selection}

In this case study we use the first seven layers from Table~\ref{table_methods}: Visual Similarity, Same Audio, Partial Audio, Music ID, Video Description, URL, and Hashtag Sequence. Audio transcripts, used to construct the two Audio layers, were available for 5,690 ($\approx 23.3$\%) videos through the Research API. The Hashtag layer could in principle be constructed, but it is not used in this study, as individual hashtags (especially popular ones)
provide a weak sign of coordination potentially leading to many false positives.

Table~\ref{table_methods} mentions additional layers that can in principle be constructed if the corresponding data is available. However, for the last five layers the corresponding information was not directly available from the Research API at the time of our data collection. 
For instance, as noted earlier, while TikTok's API grants access to video comments, it does not allow retrieval of comments linked to specific users. Engagement metrics face similar constraints: although likes and reposts can be obtained via user information, likes are rarely shared publicly, and collecting repost data is slow and inefficient, as the API does not support time-specific queries. 
At the same time, some of the missed co-actions (that is, those that would be captured by some of the layers that we could not construct) should be captured by other layers. For example, a stitch includes the original video, leading to a partial match of audio transcript. 

\subsubsection{Step 2: Similarity thresholds}

Two native German-speaking coders independently annotated 20 video pairs across 25 unique videos, evaluating similarity along three dimensions: visual similarity, audio similarity, and the underlying message conveyed. The coders reached full agreement in all but one case (95\% agreement).
These annotations formed the basis for establishing empirically grounded thresholds for the audio and video layers described below.

\begin{figure}
     \centering
     \begin{subfigure}[b]{0.45\textwidth}
         \centering
        \includegraphics[height=3cm]{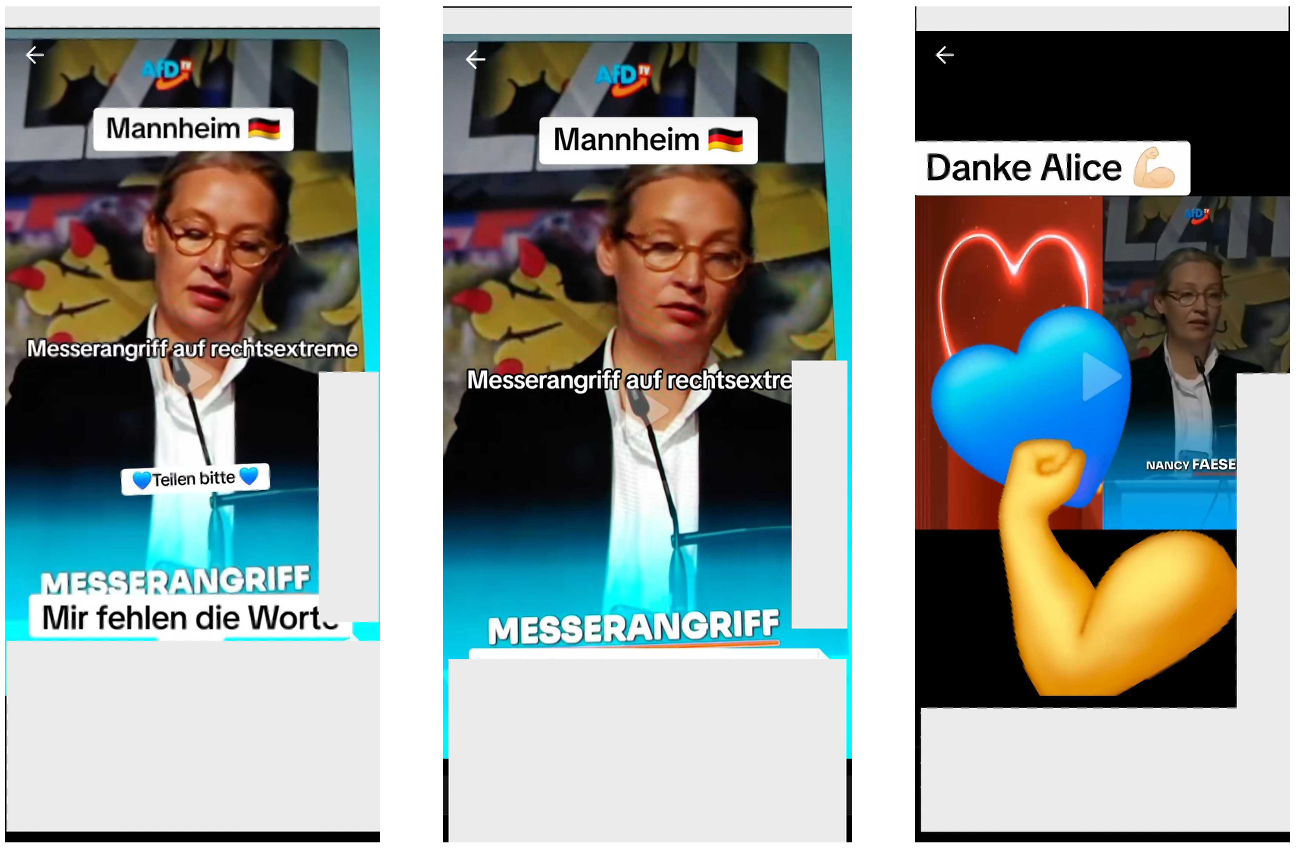}
         \caption{}
         \label{fig:ex-audio:a}
     \end{subfigure} 
     \begin{subfigure}[b]{0.40\textwidth}
         \centering
         \includegraphics[height=3cm]{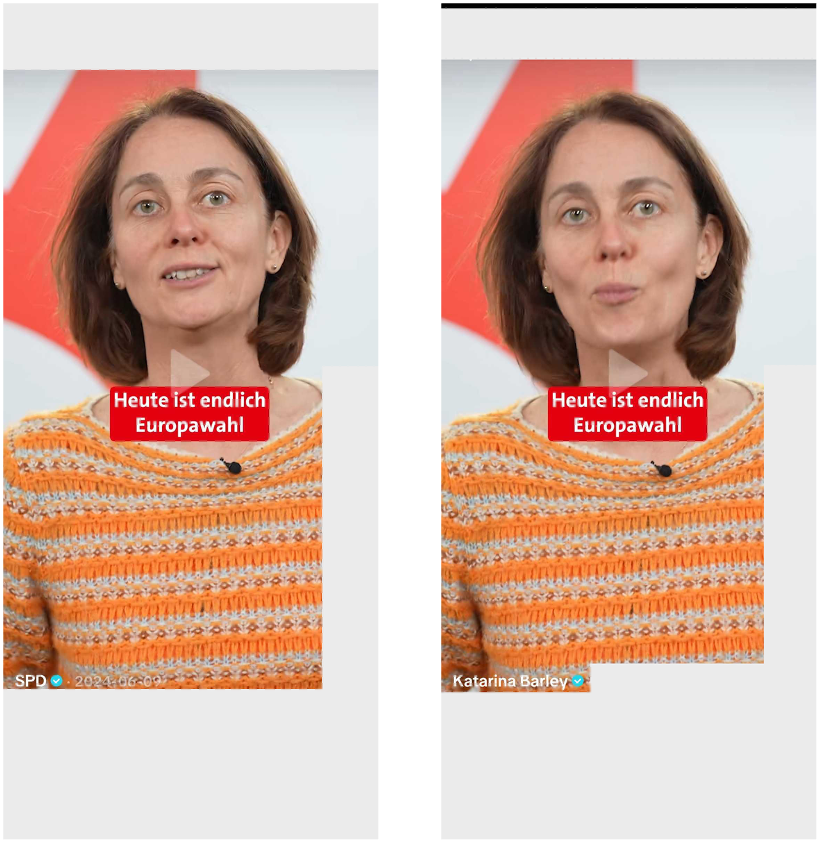}
         \caption{}
         \label{fig:ex-audio:b}
     \end{subfigure}
        \caption{Examples of videos that are considered equivalent in the Audio Layer, thanks to our fine-tuned similarity function}
        \label{fig:ex-audio}
\end{figure}

For the audio layers, we identified optimal similarity thresholds for the edit-distance metrics based on precision-recall curves: 88 for exact audio and 68 for partial matches. Both thresholds achieved perfect precision and recall in our test set. In cases where the precision-recall curve was optimal for multiple thresholds, we selected the mean of all the optimal threshold values. 
These funtions allowed us to detect duets, identical videos, and small modifications of the same video that are uploaded independently, resulting as different videos in the TikTok API.
For example, our similarity measure matches the three videos in Figures~\ref{fig:ex-audio:a} and the two videos in 
\ref{fig:ex-audio:b}. 

For visual similarity, our approach does not yield perfect precision and recall under a single threshold. The method performed well in identifying reused video snippets, including different cuts from the same interview, but is more sensitive to visual modifications such as zooming or autofill effects. Therefore, we prioritised high precision over high recall to avoid false positives, setting a similarity threshold of 1 and resulting in only detecting complete video duplicates in our test data set of 20 video pairs.

\begin{table*}[t]
\centering

\begin{tabular}{l|l|l|l|l|l|l|l}
\toprule
 &
VD &
HS &
U & MI & SA & PA & VS \\
\toprule

Nodes: 
& 554
& 1 240
& 13
& 5 049
& 186
& 480
& 425
\\

Edges: 
& 6 767
& 54 574
& 8
& 435 257
& 303
& 1 315
& 400
\\

Components: 
& 78
& 139
& 6
& 153
& 52
& 32
& 46
\\

Giant Comp. (\%): 
& 51.81
& 68.71
& 23.08
& 91.90 
& 13.98
& 84.79
& 75.29
\\

Diameter:
& 8
& 7
& 1
& 9
& 9
& 10
& 6
\\

Clustering Coeff.:
& 0.73
& 0.76
& 0.23
& 0.80 
& 0.40
& 0.41
& 0.09
\\

Density:
& 0.04
& 0.07
& 0.10
& 0.03
& 0.02
& 0.01
& 0.00 
\\ \bottomrule

\end{tabular}
\caption{Structure of the individual unfiltered layers, VD= Video Description, HS= Hashtag-Sequence, U= URL, MI= Music ID, SA = Same Audio, PA= Partial Audio, VS= Visual Similarity.}
\label{table1}
\end{table*}

\subsubsection{Step 3: Frequency and temporal filtering}

The previous two steps generated seven layers, each corresponding to a co-action network, summarised in Table \ref{table1}.
The seven unfiltered networks have significantly different sizes.
About 50\% of the users present in our dataset use the same Music ID as another user at least once. On the other side, the smaller audio transcript networks are most likely due to the audio transcript not being available for many posts, e.g. slide shows. The third step in our method is to identify thresholds for the frequency and time-based filters, so that some of the edges in the dense layers likely not due to coordination are removed before the detection of coordinated groups of accounts.

\begin{table*}[h!]
\small
\centering
\begin{tabular}{ll}
\toprule
\textbf{Filter} & \textbf{Type of threshold (\emph{value})} \\
\toprule
Hashtag-Sequence & Frequency (\emph{10 times}) \\ \midrule
Video Description & Frequency (\emph{10 times}) \\ \midrule
URLs & No Filtering \\ \midrule
Music ID & Frequency (\emph{Above Avg.}) \\ \midrule
Partial and same audio &  No Filtering \\ \midrule
Visual Similarity & No Filtering \\
\bottomrule
\end{tabular}
\caption{Chosen threshold used to filter the edges on each layer.}
\label{tab:filtering}
\normalsize
\end{table*}

In the frequency-based filters, we retain edges only if the number of shared co-actions between a pair of users (i.e.~edge weight) was at least 2, 10, or equal to the average edge weight for that layer. These values are chosen based on previously utilised thresholds in the literature~\parencite{Soares2023, Schoch2022, Ng2023_b}.
  
In the time-based filters, we retain only those edges for which the annotated time difference between co-actions was equal to or less than 60 seconds, 120 seconds, or 5 minutes. These thresholds also follow precedents in prior work~\parencite{Soares2023, Ng2023_b, giglietto_tiktok_viz}.

\begin{figure}
     \centering
     \begin{subfigure}[b]{0.3\textwidth}
         \centering
        \includegraphics[width=\textwidth]{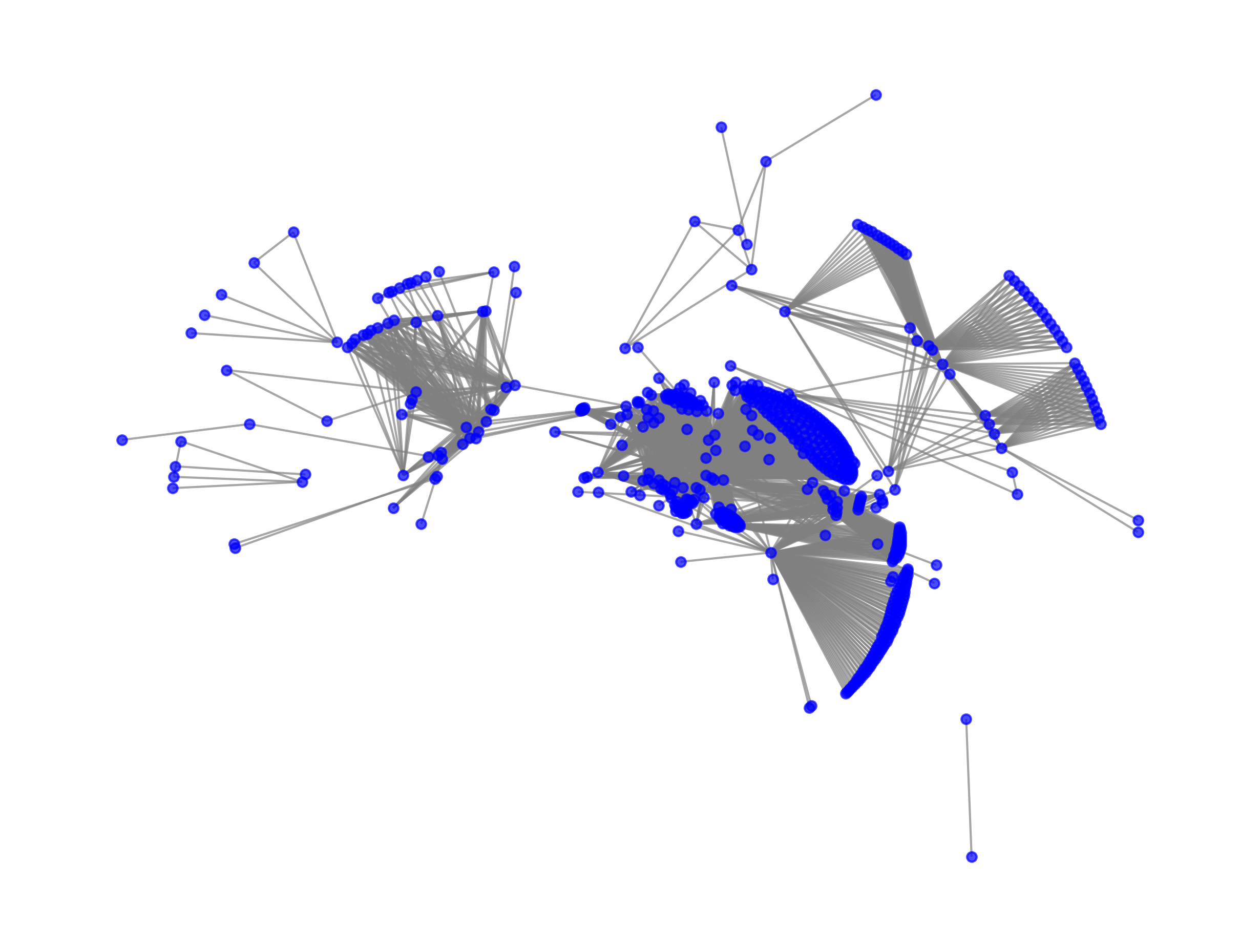}
         \caption{Hashtag Sequence}
         \label{fig:network_hashtags}
     \end{subfigure}
    \centering
    \begin{subfigure}[b]{0.3\textwidth}
        \includegraphics[width=\textwidth]{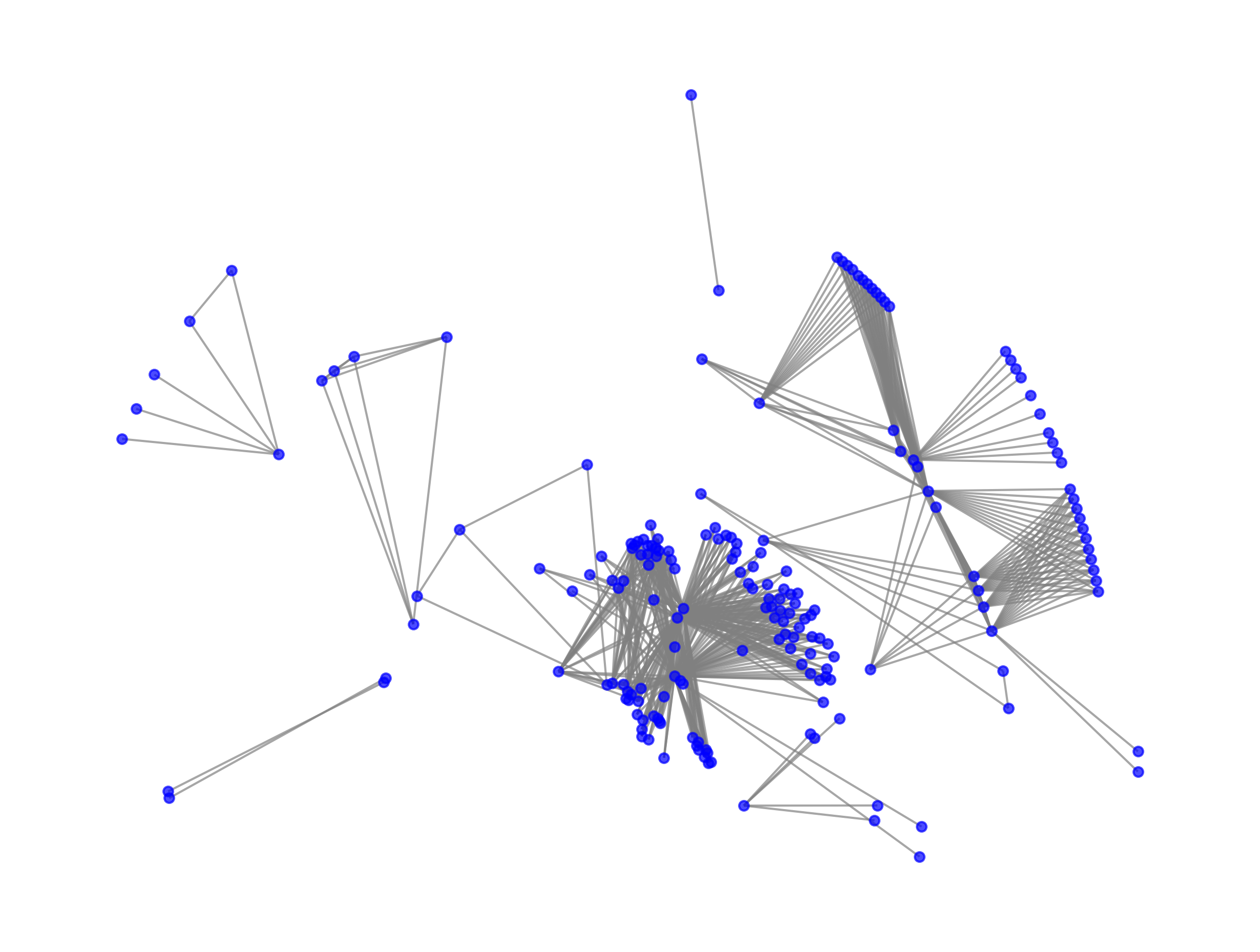}
        \caption{Video Description}
        \label{fig:ex-videodescription}
    \end{subfigure}
        \begin{subfigure}[b]{0.3\textwidth}
        \includegraphics[width=\textwidth]{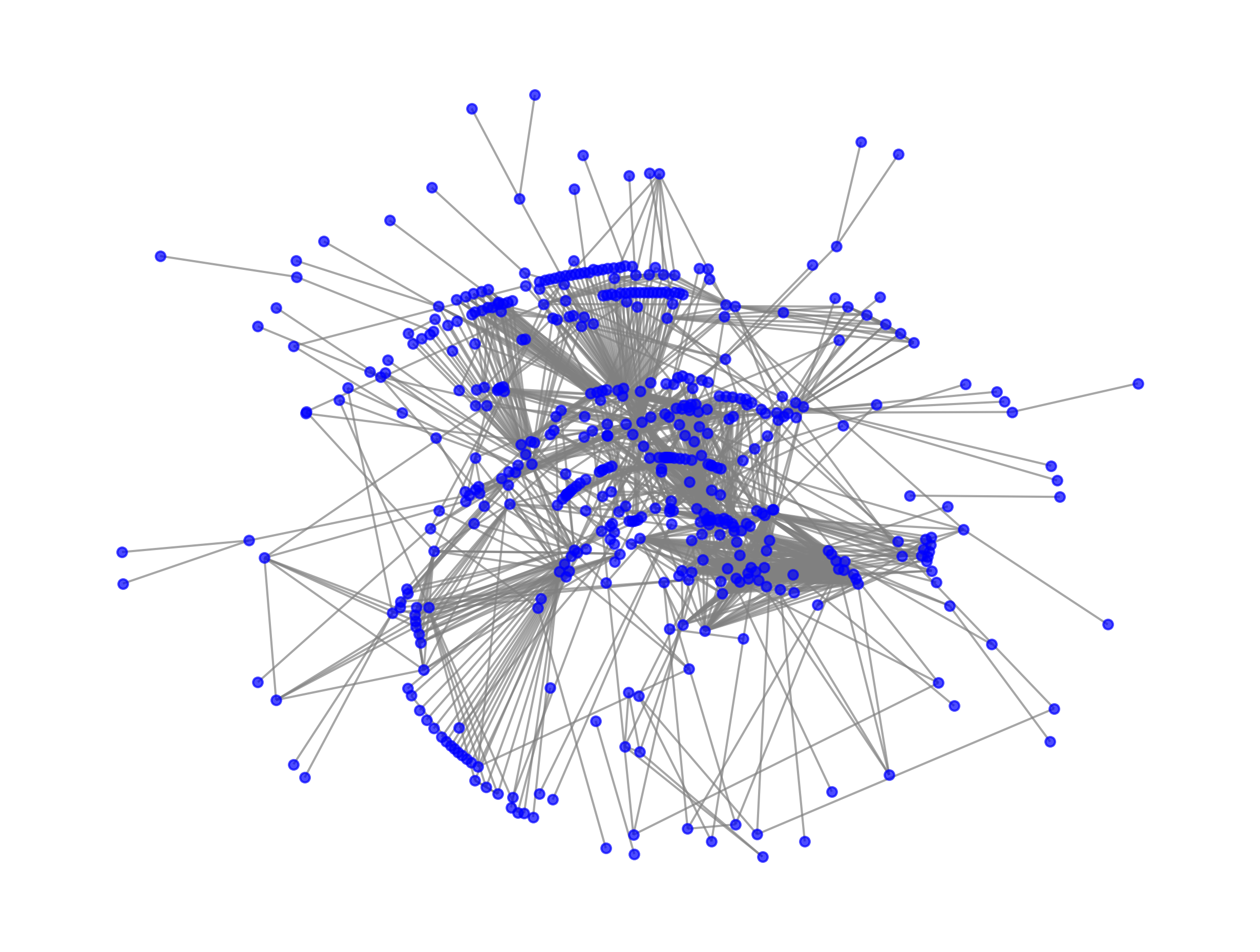}
        \caption{Partial Audio}
        \label{fig:ex-partialaudio}
    \end{subfigure}
        \begin{subfigure}[b]{0.3\textwidth}
        \includegraphics[width=\textwidth]{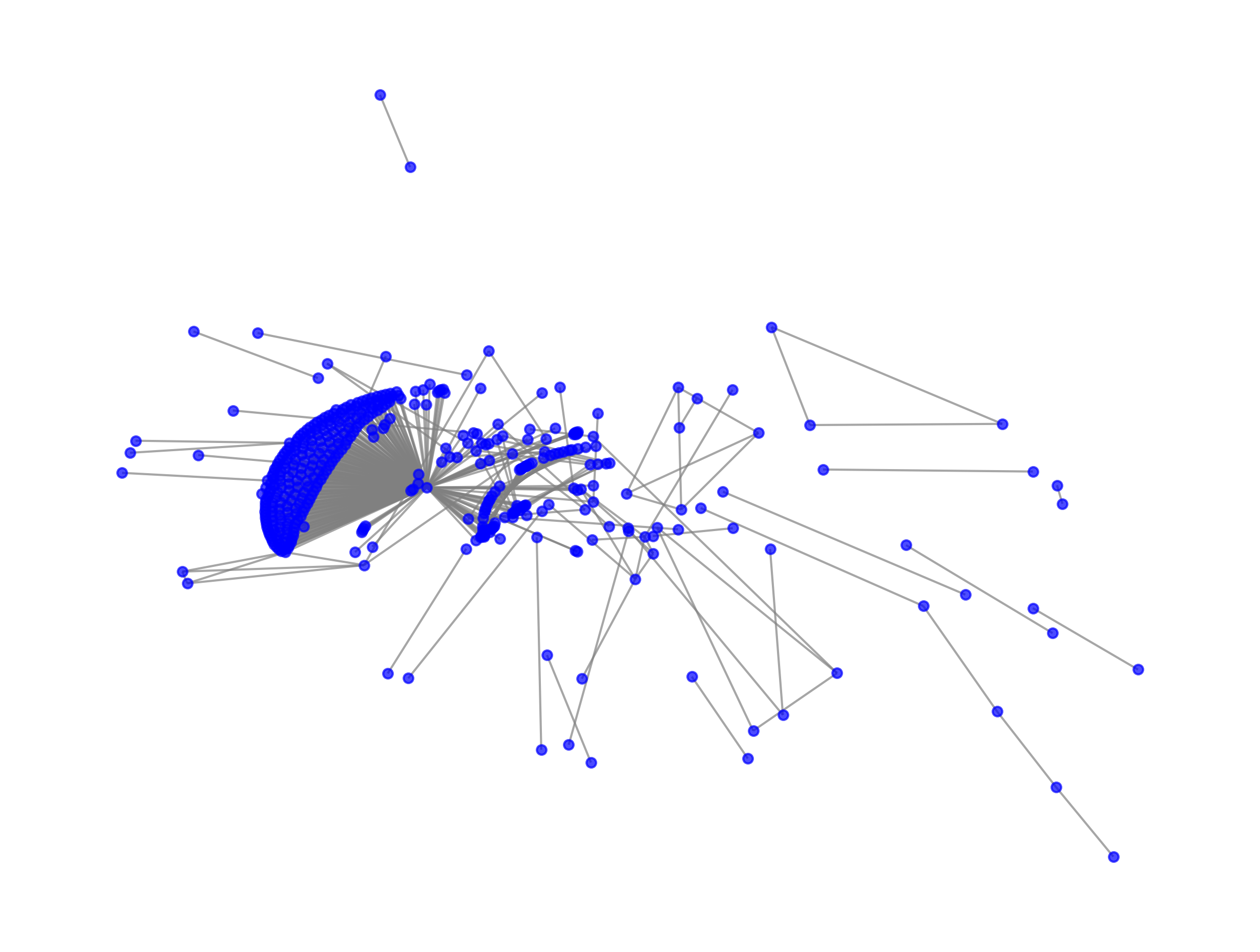}
        \caption{Visual Similarity}
        \label{fig:ex-videosimilarity}
    \end{subfigure}
        \begin{subfigure}[b]{0.3\textwidth}
        \includegraphics[width=\textwidth]{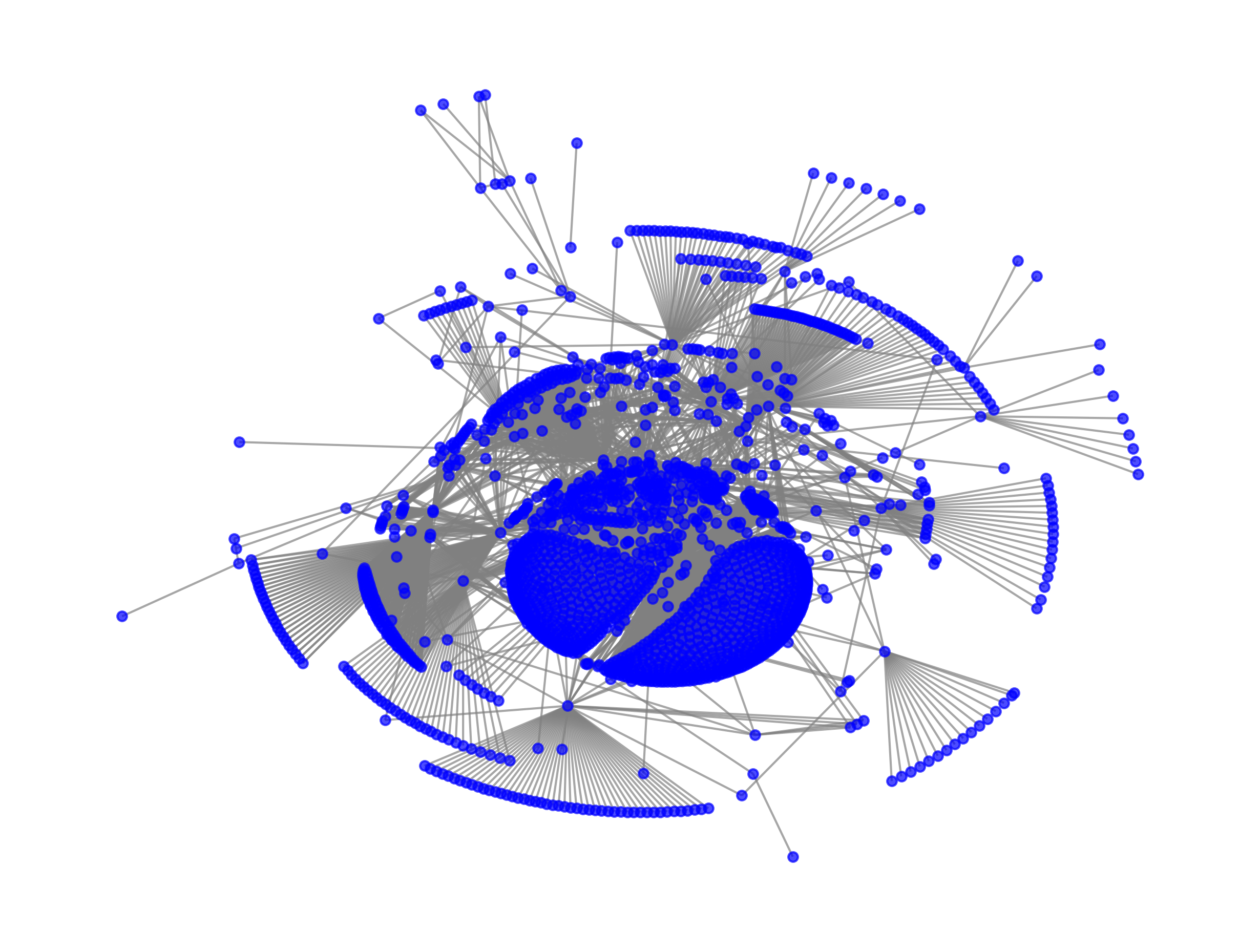}
        \caption{Music ID}
        \label{fig:ex-musicid}
    \end{subfigure}
            \begin{subfigure}[b]{0.3\textwidth}
        \includegraphics[width=\textwidth]{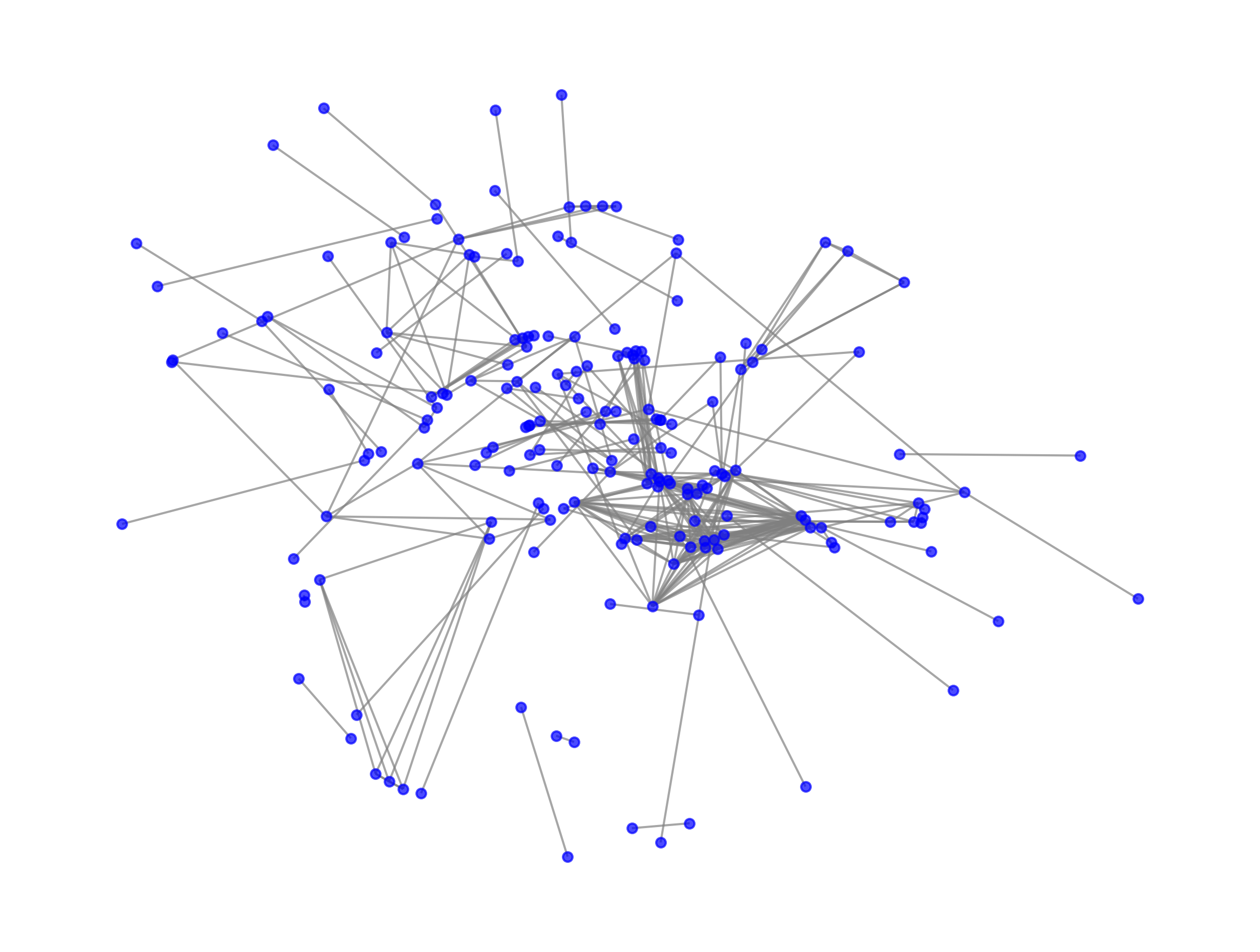}
        \caption{Same Audio}
        \label{fig:ex-sameaudio}
    \end{subfigure}
        \caption{Visualisation of the different filtered layers of the network}
        \label{fig:networkvisualisation}
\end{figure}

\begin{table*}[t]
\centering

\begin{tabular}{l|l|l|l|l|l|l|l}
\toprule
 &
VD &
HS &
U & MI & SA & PA & VS \\
\toprule

Nodes: 
& 190
& 568
& 13
& 1 877
& 186
& 480
& 425
\\

Edges: 
& 552
& 2 668
& 8
& 9 588
& 303
& 1 315
& 400
\\

Components: 
& 14
& 18
& 6
& 12
& 52
& 32
& 46
\\

Giant Comp. (\%): 
& 51.58
& 89.96
& 23.08
& 96.91 
& 13.98
& 84.79
& 75.29
\\

Diameter:
& 2
& 10
& 1
& 9
& 9
& 10
& 6
\\

Clustering Coeff.:
& 0.72
& 0.70
& 0.23
& 0.75 
& 0.40
& 0.41
& 0.09
\\

Density:
& 0.03
& 0.02
& 0.10
& 0.01
& 0.02
& 0.01
& 0.00 
\\ \bottomrule

\end{tabular}
\caption{Structure of the individual filtered layers, VD= Video Description, HS= Hashtag-Sequence, U= URL, MI= Music ID, SA = Same Audio, PA= Partial Audio, VS= Visual Similarity.}
\label{table1_filtered}
\end{table*}

This filtering process yielded a set of refined co-action layers, shown in Figure \ref{fig:networkvisualisation}, each capturing stronger and more plausible signs of coordination. Notice that the temporal proximity of interactions has not been as informative as the frequency of such interactions (See Table~\ref{tab:filtering}). We find that often similar content is posted over a longer period of time. 
In addition, co-actions which are common in the data, are likely to appear within a short time-frame by chance. These layers formed the basis for the structural and content-based analysis performed in the following steps. 

\subsubsection{Step 4: Multilayer clustering}

\begin{table*}[t]
\centering

\begin{tabular}{l|llll}
\toprule
    & CP (8,1) & CP (4,2) & CP (3,3) & GLouvain \\
    \toprule
    Number of clusters    &  17  & 16 & 1     & 86 \\ \midrule
    Avg. users/cluster  &  33.94  & 13.06 & 3.00  & 34.66  \\ \midrule
    Proportion of users in clusters      &  0.15  & 0.11 & 0.06      & 1.00  \\ \midrule
    Avg. layers/cluster    &  1.12 & 2.00 &   3.00        & 2.00 \\ \bottomrule
\end{tabular}
    \captionof{table}{Summary of the four clusterings computed on our data}
    \label{tab:commdetection}
\end{table*}

We executed two types of multilayer clustering algorithms. Multilayer clique percolation (CP) has two parameters to control the strength of the clusters, both in term of the minimum number of accounts ($k$) that must all be adjacent (clique) to be included in a cluster, and the minimum number of layers ($m$) where a clique must be present to be included in a cluster. This can be considered an additional filter, as less accounts are included in the clusters for higher values of the parameters. We use three parameter configurations: $k=8, m=1$, notated CP(8,1), focusing on very strong clusters appearing on at least one layer, and CP(4,2) and CP(3,3), focusing on cross-layer clusters. We also use generalised Louvain (GLouvain), the algorithm that is normally applied in the literature on network-based coordination detection, which includes all accounts in clusters.
To execute the clustering algorithms, we used the multinet library \parencite{Magnani2021}.  
Table \ref{tab:commdetection} summarises the resulting clusterings for our four configurations. As expected, GLouvain produces many large clusters. CP(8,1) only retains 15\% of the accounts, grouped into 17 clusters, with most clusters being present on a single layer. Increasing $m$, as expected, produces fewer and smaller clusters spanning multiple layers, down to a single cluster of three users all co-acting on three different layers for CP(3,3).

\subsubsection{Step 5: Cluster characterisation}

In this section we present a systematic analysis of the 17 clusters returned by CP(8,1), summarised in Table \ref{tab:commsystematicreview}. For space reasons, we do not present detailed cluster information for all four clusterings, but we provide some examples of clusters from the three other clusterings at the end of this section. 

\begin{table*}[p] 
\centering

\begin{tabular}{l|llllll}
\toprule
      & n  & Layers & Similar & Sync. & False  & User overlap  \\
         & (users) & & Names &  Posting & Positives & \\
    \toprule
    0    &  10 & PA & no & once&  & 9,14,16  \\ \midrule
    1  &  9 & PA & no & no & & \\ \midrule
    2   &  8 & HS & some & no& & 4 \\ \midrule
    3    &  16 & PA,SA & no & some & & 7,8,12,13,16 \\ \midrule
    4  &  8  & HS & some & some& & 2 \\ \midrule
    5   &  8  & MI & no & once& & 8,13,16 \\ \midrule
    6   &  15 & PA & some & some & & 16\\ \midrule
    7  &  76  & MI & no & some & yes   & 3,10,12,16\\ \midrule
    8   &  55 & HS & no & some& & 3,5,9,12,13,16\\ \midrule
    9    &  46 & HS & no & some& & 0,8,13,16\\ \midrule
    10  &  22  & HS & no & some& yes  & 7 \\ \midrule
    11   &  9  & PA & no & once & & \\ \midrule
    12    &  29 & PA & some & some& & 3,7,8,13,16 \\ \midrule
    13  &  13 & HS,VD & no & once & & 3,5,8,9,12,16 \\ \midrule
    14   &  12 & PA & no & once& & 0,16 \\ \midrule
    15   &  11 & PA & some & no& & 16 \\ \midrule
    16   &  230 &  MI & some & some& & 0,3,5,6,7,8,9 \\
       &   &   &  & & & 12,13,14,15\\ \bottomrule
\end{tabular}
    \captionof{table}{Summary of the clusters identified using CP (8,1). n = user count, Layers = Layer cluster appears on, Similar Names = if users appear to have similar usernames within the cluster, Sync. Posting = if users post within short (60 s) timeslots of each other, False Positives = if there is clear false positives, such as different parties represented within the same cluster, User Overlap = in which other clusters we find some of the same users}
    \label{tab:commsystematicreview}
\end{table*}

Let us first look at the number of users and the layers where each cluster is present, indicated in the first columns of Table \ref{tab:commsystematicreview}. We see that some layers do not contain any clusters.
This does not imply that coordination is absent from these layers: for example, the Visual Similarity layer (where no clusters are found) does include pairs of accounts that posted the exact same video, such as party accounts reposting the video of their candidate or vice versa. However, CP(8,1) only focuses on large coordinated groups, with at least 8 accounts. Other layers, such as Partial Audio, Music ID and Hashtag Sequence, include multiple clusters. Video Description, being a subset of Hashtag Sequence, only appears in clusters across these two layers. Looking at the results of the clustering algorithm, it is apparent that each layer includes unique clusters, only determined by the corresponding modality and similarity function, while the same users are sometimes present in multiple clusters across different layers. 

\begin{figure}
     \centering
     \begin{subfigure}{0.325\textwidth}
         \centering
         \includegraphics[height=2.5cm]{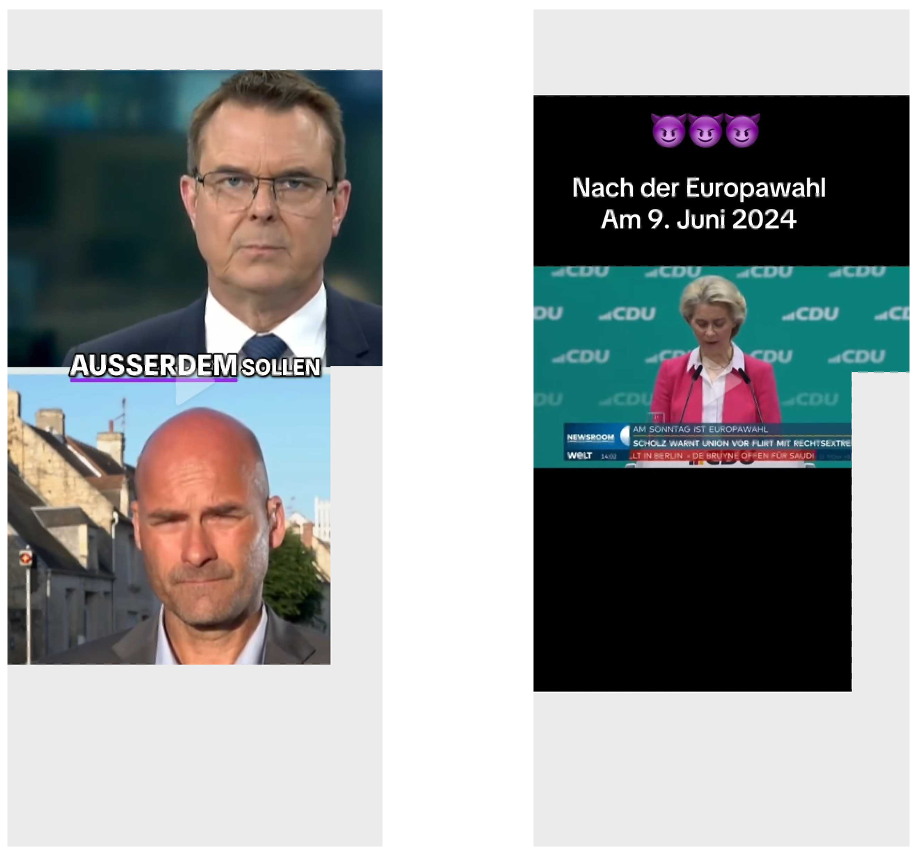}
         \caption{}
         \label{fig:ex-audio:c}
     \end{subfigure}
     \begin{subfigure}{0.325\textwidth}
         \centering\includegraphics[height=2.5cm]{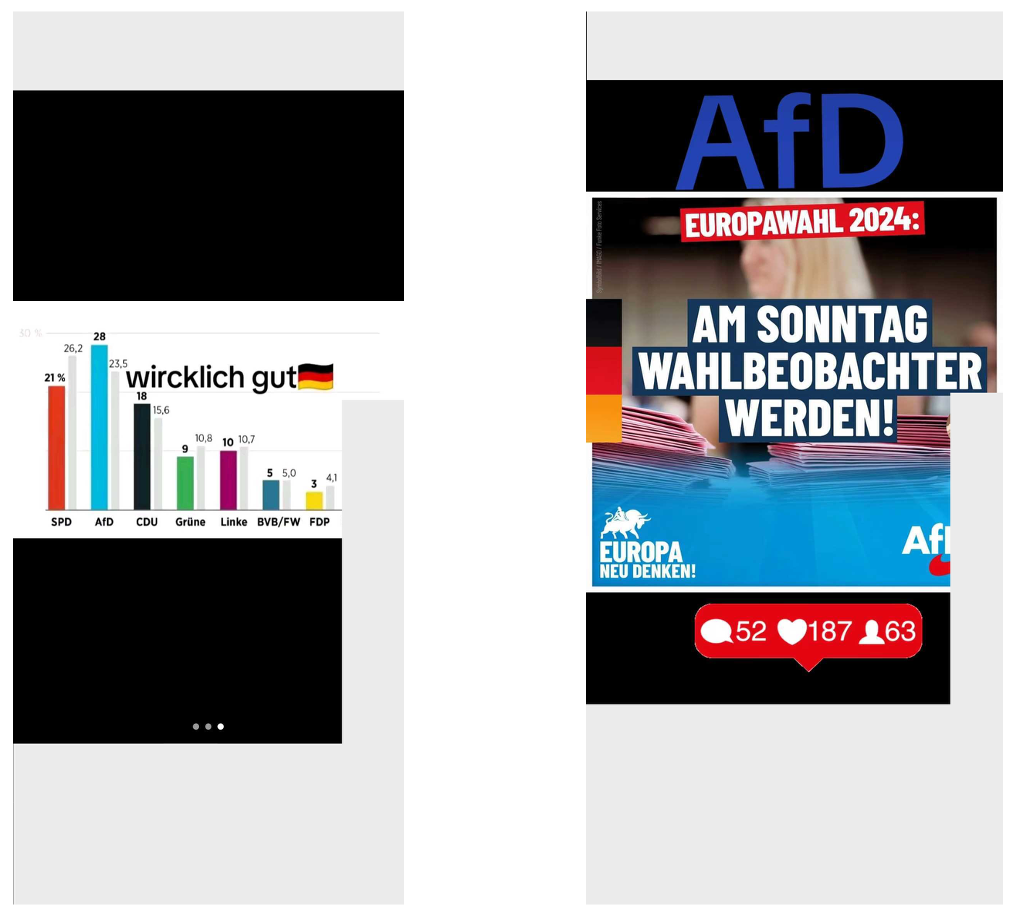}
         \caption{}
         \label{fig:ex-music:a}
     \end{subfigure}
    \begin{subfigure}{0.325\textwidth}
         \centering
        \includegraphics[height=2.5cm]{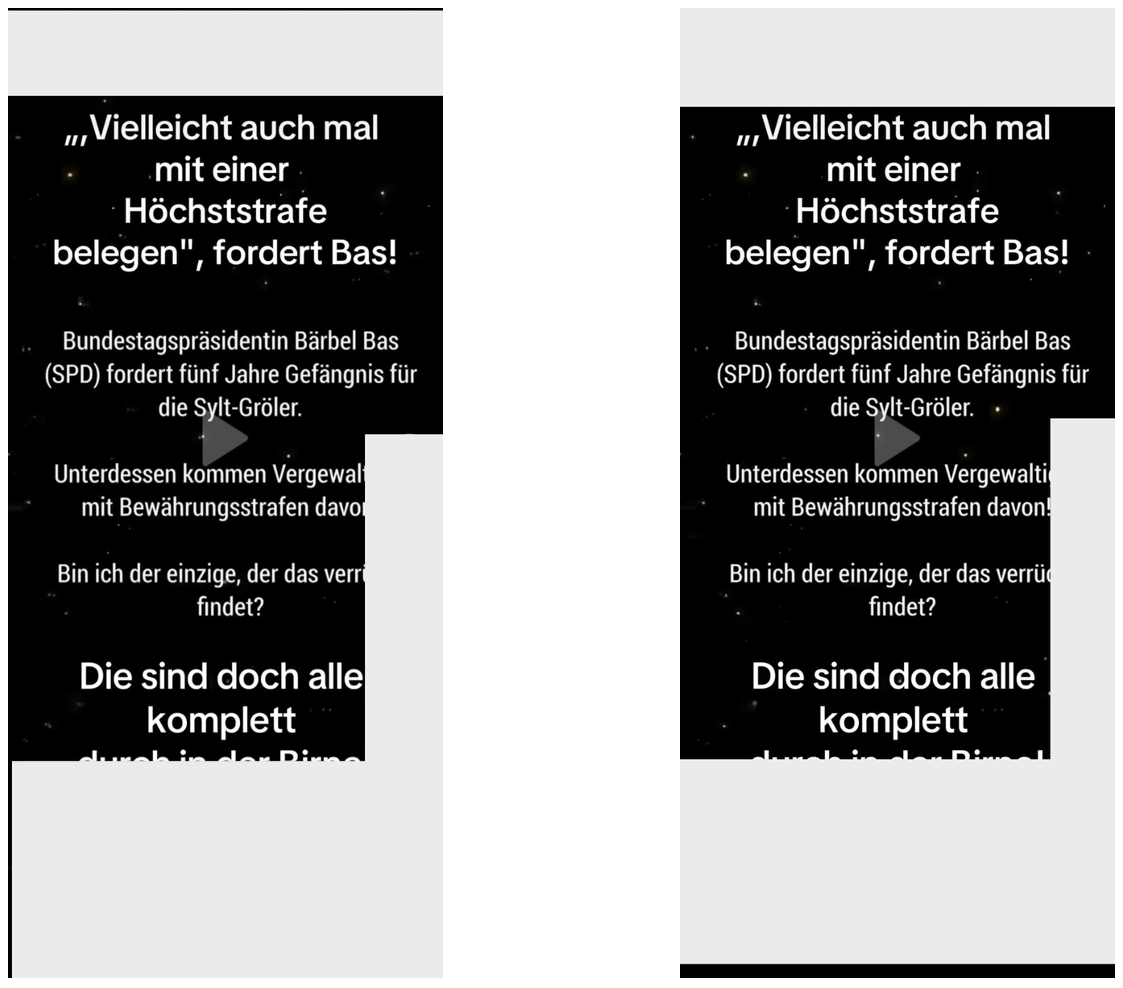}
        \caption{}
        \label{fig:ex-hs}
    \end{subfigure}
        \caption{Examples of pairs of videos found in the same clusters using clique percolation}
        \label{fig:ex-music}
\end{figure}

When we look in more detail at the accounts and content characterising the clusters, we observe clusters with different characteristics on different layers. Partial Audio, for example, contains clusters where some accounts are using terms like ``aktuell'' and ``nachrichten'' to appear as if real news agency might be behind the accounts. Some of them were only active during the period leading up to the election (Figure \ref{fig:ex-audio:c}, Cluster 3 and 11 in Table \ref{tab:commsystematicreview}). 

Music ID contains the largest clusters, with different characteristics compared to the other layers. As an example, we find one large cluster where a majority of the connections are based on using the song ``L’amour Toujours'' as their audio (Figure~\ref{fig:ex-music:a}, Cluster 16 in Table \ref{tab:commsystematicreview}). This song has been known for its popularity within the political right wing in Germany and caught nation-wide attention when a video appeared of people at a bar shouting racist sentences to the song on the 19th of May, in the lead up to the European election, causing a nation wide backlash. The cluster contains videos with the same underlying message in support of the right-wing party. These accounts are still active today, and despite the easy-to-produce content of the post, they do not show any clear signs of automation. Therefore, this case might rather show the power of using sound as a collective, where sound creates a social bond that unites the supporters of the right-wing party. This phenomenon has been discussed in previous research \parencite{Bsch2024}. 

The Hashtag Sequence layer reveals clusters of potential coordinated accounts. Among the users in these clusters, we find user accounts' names starting with the word ``user" followed by a set of numbers, suggesting automated account creation. The videos are usually video snippets of news or talk shows with comments written within the videos. Often this text contains a bias towards the politicians or content appearing in the videos. Most of the user accounts in these clusters were only active around the election time and have since stopped posting. As an example, Figure \ref{fig:ex-hs} (Cluster 9 in Table \ref{tab:commsystematicreview}) shows two videos from a cluster of similar user account names. 

An interesting observation is that we  find user accounts appearing in multiple clusters and on different layers (See last column of Table \ref{tab:commsystematicreview}, showing how different clusters overlap with others). Specifically, user accounts of a likely coordinated cluster on the Video Description and Hashtag Sequence layers appear to engage with popular audio tracks used by other subgroups (Cluster 16 in Table \ref{tab:commsystematicreview}). This highlights how coordinated actors may leverage different modalities, such as shared audio content, to establish links with otherwise separate groups and amplify their reach. Especially the use of a popular audio track (Music ID) should increase the reach, while potentially being masked by a high amount of organic activity in the same vein. We also find clusters that could be considered part of a joint coordinated effort, but because of small differences in the actions of their users they do not appear as a single cluster (Clusters 2 and 4 in Table \ref{tab:commsystematicreview}).

\begin{figure}
     \centering
     \begin{subfigure}[b]{0.6\textwidth}
         \centering
        \includegraphics[height=2.5cm]{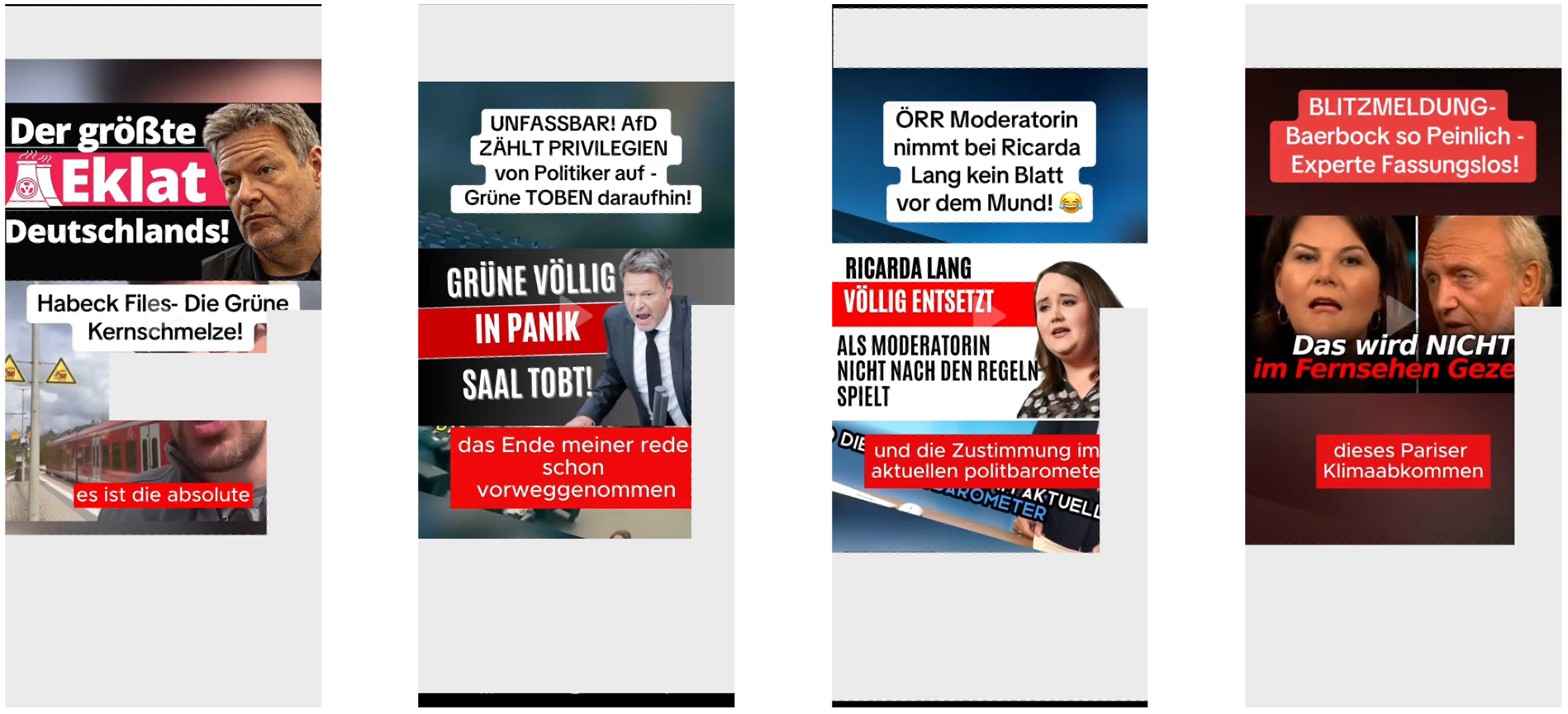}
         \caption{}
         \label{fig:ex-vd:a}
     \end{subfigure}
     \begin{subfigure}[b]{0.3\textwidth}
         \centering
         \includegraphics[height=2.5cm]{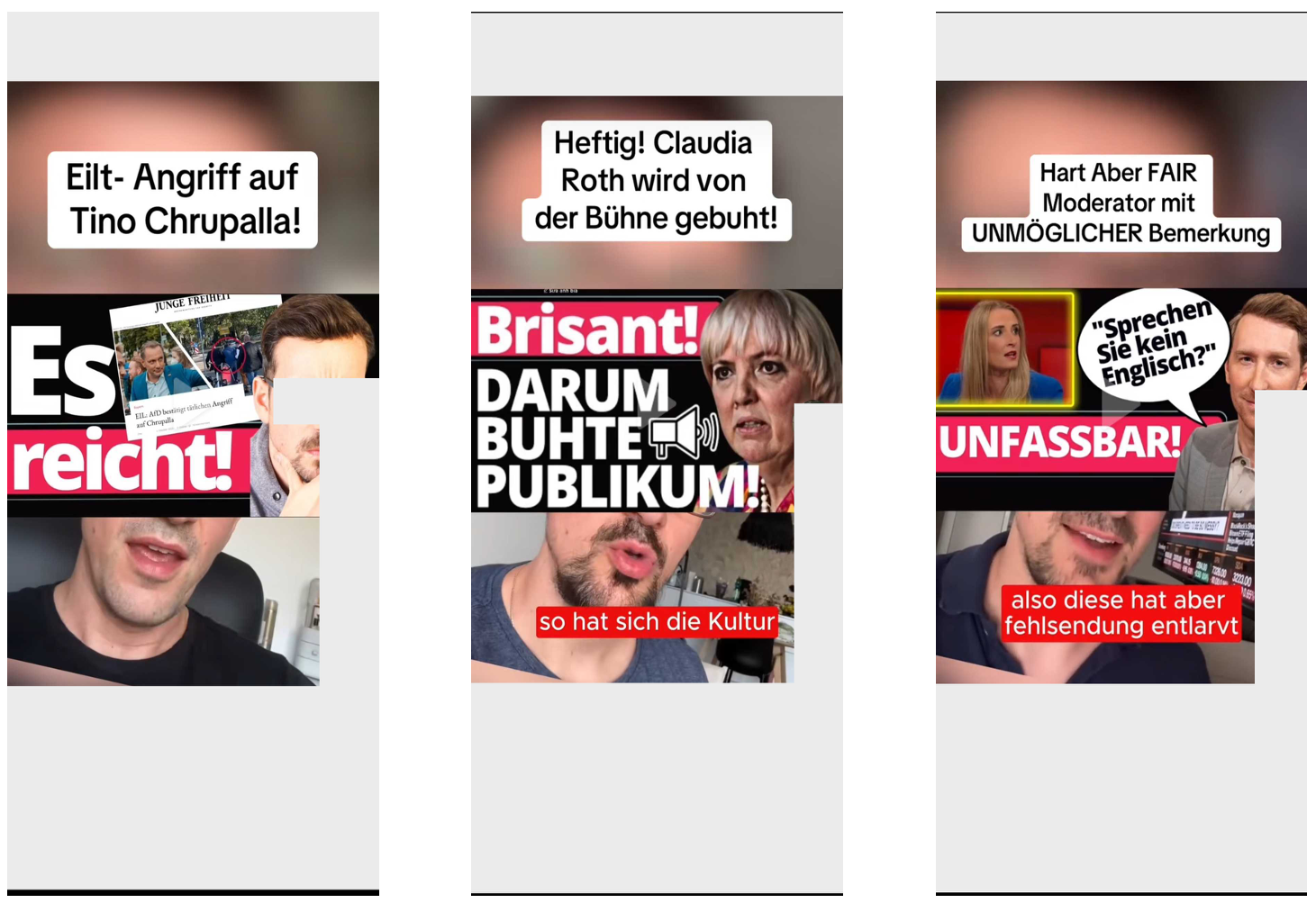}
         \caption{}
         \label{fig:ex-vd:b}
     \end{subfigure}
        \caption{Examples of videos from two clusters found in the Video Description and Hashtag Sequence layers using CP(4,2)}
        \label{fig:ex-vd}
\end{figure}

We conclude by providing a few examples from the other clusterings, using different parameters for clique percolation and using GLouvain. With CP(4,2) we allow the algorithm to find smaller clusters, but we only focus on clusters present on at least two layers. One such cluster consists of accounts where all usernames are starting with the same word (Figure~\ref{fig:ex-vd:a}). Another cluster has usernames typically starting with a female name and ending with a set of numbers (Figure~\ref{fig:ex-vd:b}). These accounts post clips from talkshows or news shows, while adding a bias through additional text in the video towards the politicians appearing in the clips. Some also include videos in ``vlogging'' style, making the account appear more authentic. However, the same person appears on multiple accounts. Most accounts were only active specifically around the European election and have since not posted. 
Using CP (3,3), we only found one cluster included in three layers, containing three accounts (party accounts and the main candidate). 

\begin{figure}
     \centering
     
     \begin{subfigure}[b]{0.45\textwidth}
        \centering\includegraphics[height=3cm]{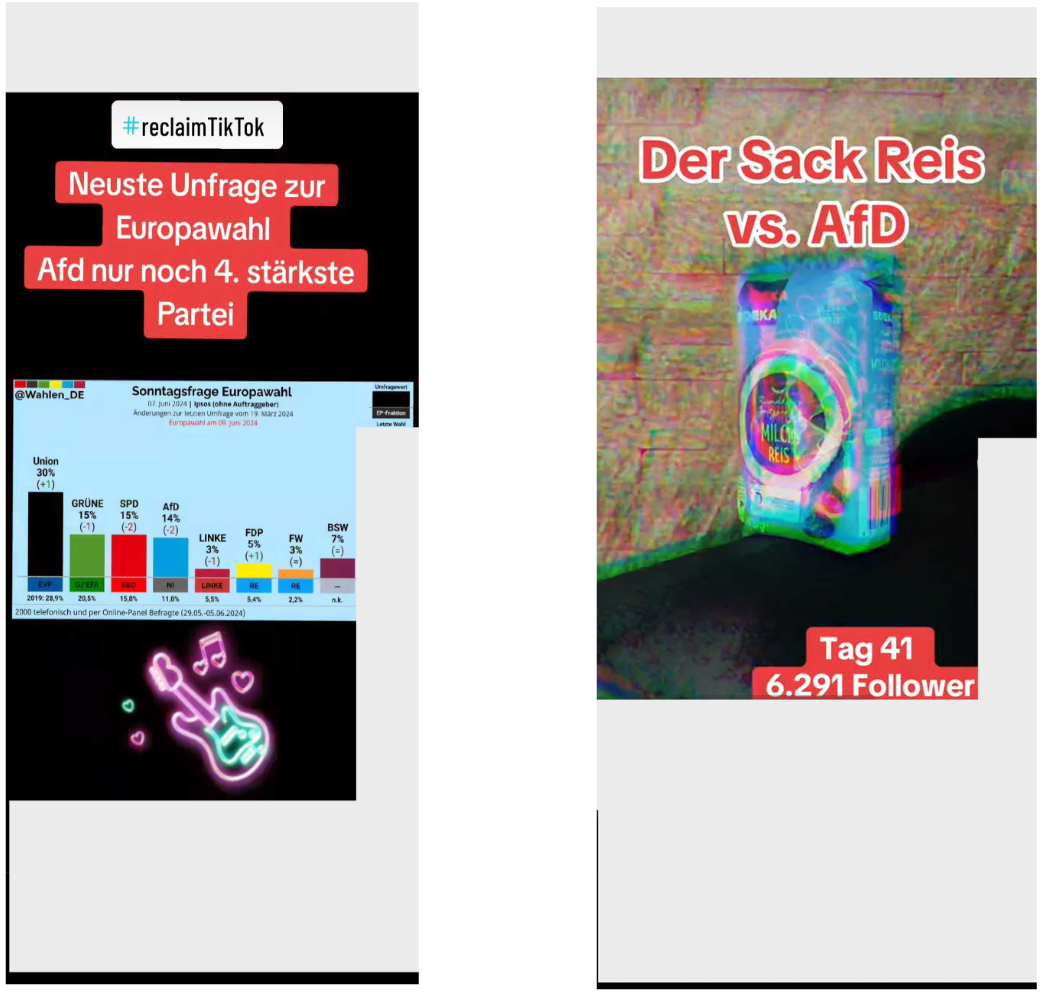}
         \caption{}
         \label{fig:ex-music:b}
     \end{subfigure}
    \begin{subfigure}[b]{0.5\textwidth}
        \centering
        \includegraphics[height=3cm]{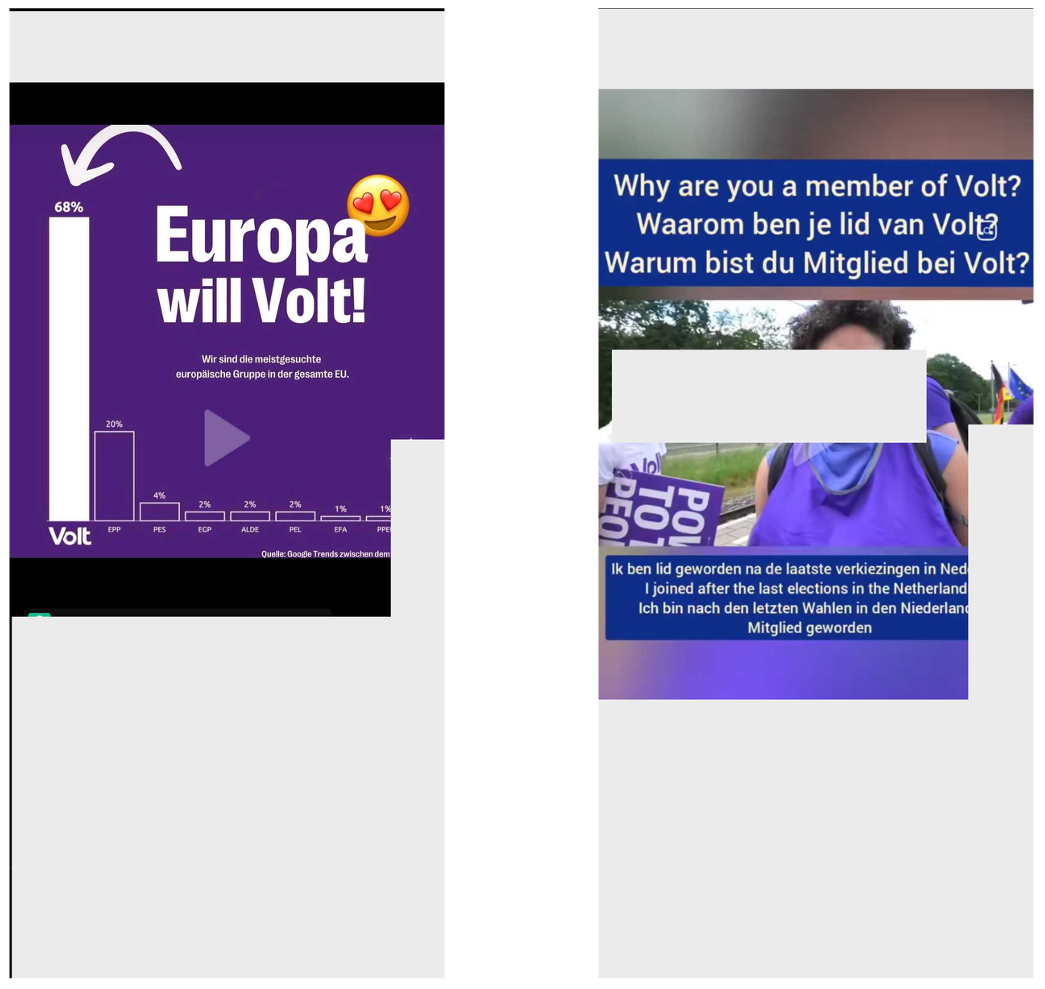}
        \caption{}
        \label{fig:ex-hs:a}
    \end{subfigure}
        \caption{Examples of videos from two clusters found with GLouvain}
        \label{fig:ex-music}
\end{figure}

Some clusters are only detected using GLouvain, as it is the least strict out of the clustering algorithms we ran. For example, we find a smaller cluster in the Music ID layer, generated by the usage of audio attributed to the political left wing, such as the audio track ``N*zis raus" (Figure~\ref{fig:ex-music:b}). Interestingly, while the cluster promotes a different message, the type of content resembles the content type of the ``L'amour Toujours" cluster. Accounts post easy-to-produce content, as for example a screenshot with added text and music. Nonetheless, user accounts remain active until now. Further, the cluster uses the ``Reclaim TikTok'' hashtag, which is a movement started in the lead up to the election.
While this is a political hashtag, it is not directly associated with one party, which might be the reason for an under-representation of the movement in our dataset. Furthermore, we find clusters containing the party campaigns from the different party accounts based on the video description, such as in the example in Figure \ref{fig:ex-hs:a} from the party Volt.

\section{Discussion and Conclusion}
This paper presents an approach for detecting coordinated behaviour on video-first platforms that addresses the challenging combination of multimodality and complex similarity characterising these online environments. Our approach uses multilayer network analysis as a way to identify signs of coordination within and across modalities, and fine-tuned similarity functions to handle similar but not identical content carrying the same message. We focus on TikTok, as a pre-eminent video-first platform, and, empirically, on the 2024 European Parliament elections in Germany.

Our empirical case demonstrates the capacity of our approach to detect different types of coordinated behaviour. Like \textcite{Luceri2025}, who focused on US election data, we find clusters of user accounts with similar user names, and clusters of accounts that were only active during the election period, suggesting that these accounts may have created for the specific purpose of amplifying political narratives during the studied event. Given the algorithmic design of the platform,  it is possible to get a large number of views without followers, which in turn means it is possible for custom created accounts to gain reach. In summary, our approach can capture an important type of online behaviour in an increasingly important type of online environments.

\subsection{Limitations and pitfalls}

The present application of our approach is circumscribed by the current capabilities of the TikTok Research API, such that not all meaningful similarity measures are possible to examine. For example, measures such as comments, reposts and likes are not fully accessible through the API, either because they are not made available or because it is not possible to set a time frame that would reduce the number of calls to the API down to a reasonable level.  

It is also important to consider that the areas in which organic user engagement is encouraged by the platform clouds the analysis of strategically coordinated behaviour. Our analysis shows that the similarities that the platform encourages, such as using the same audio snippet (Music ID) as others, stand out as larger networks. As a consequence our similarity detection algorithms likely detect a significant proportion of organic co-actions not based on coordination. At the same time, the most popular areas highlight content styles that may prove particularly effective as focal points for strategic coordination.

We also note that while our data serves our present purposes of applying our approach, a full empirical study of coordinated behaviour on TikTok in the German context of the European Parliament elections is out of the scope of this paper. Our case study delivers, at best, a partial view of communication around the elections, in part because of the character of the data. There was a time lapse between the election and the collection of the data, as a consequence posts may have been deleted in the interim. In addition, we focused specifically on hashtags related to the different parties, which excluded non-party actors, topics or hashtags, such as for example the ``Reclaim TikTok'' hashtag. 

While our method finds clusters indicating high likelihood of coordination, some clusters are clear false positives. For example, some clusters contain user accounts of highly different political parties or political organisations with opposing views. One of the reasons for these false positives is that some posts use 
party-related hashtags (also used by the parties themselves)
to formulate criticism, or
general hashtags referring to the elections. While our filters remove some of these cases, others are still present in the final clusters especially when a complete clustering algorithm (assigning all accounts to a cluster) is used.
Furthermore, although we find small clusters in which all user account activity was concentrated in the period leading up to the European elections, larger clusters are generally a mixture of accounts that continue to actively post and those whose activity was concentrated during the elections.

At the same time, there is a risk of false negatives, that are generated by the technical design of the platform, direct content manipulation, or limitations in the data that is made available through the API. 
Platform conditions are also a source of potential false negatives. While false positives can be filtered out when the nature of the clusters is established, in the last step of our methods, false negatives are not captured by the method and are thus lost. In studies focusing on text-based platforms, one commonly used sign of coordination is when many users separately upload the same content in a short time-frame. In principle, on TikTok, we could also look for the same video being uploaded by multiple users. However, the uploading process may lead to differences in the video as it appears on the platform, not because of any explicit intention to modify the content, but because of technical format issues.
Videos on TikTok are usually in portrait format, but video snippets from TV shows or YouTube are often in landscape format. In order to fit those videos into the portrait format of TikTok, some users zoom in, while others autofill the space. These differences in data processing may lead to false negatives.

\subsection{Looking forward}

An important consideration in this emerging area is the need for thoughtful ethical reflection when engaging with visual content, social media data, and a relatively new platform for which research best practices are not well established. To this end, we have, among other things, taken extra care to anonymise user account information and personal identifiers such as faces in visuals of non-public figures. 

Our proposed methodology based on multilayer network analysis shows a promising path to address coordination on video-first platforms such as TikTok. At the same time, we identify avenues for future development. These include the design and validation of additional similarity, filtering, and clustering methods for specific modalities, and further exploration of cross-layer patterns.